\documentclass[conference]{IEEEtran}
\IEEEoverridecommandlockouts
\usepackage{cite}
\usepackage{todonotes}
\usepackage{amsmath,amssymb,amsfonts}
\usepackage{algorithmic}
\usepackage{graphicx}
\usepackage{textcomp}
\usepackage{xcolor}
\usepackage{flushend}
\usepackage{float}
\usepackage{circledsteps}

\begin{document}

\title{A Rapid Prototyping Language Workbench for Textual DSLs based on Xtext:\\ Vision and Progress
}

\author{\IEEEauthorblockN{Weixing Zhang\IEEEauthorrefmark{1}, Jan-Philipp Steghöfer\IEEEauthorrefmark{2}, Regina Hebig\IEEEauthorrefmark{3}, Daniel Strüber\IEEEauthorrefmark{1}\IEEEauthorrefmark{4}}
\IEEEauthorblockA{
\textit{\IEEEauthorrefmark{1}Chalmers $|$ University of Gothenburg, Gothenburg, SE} \ 
\textit{\IEEEauthorrefmark{2}XITASO GmbH, Augsburg, DE} \\
\textit{\IEEEauthorrefmark{3}University of Rostock, Rostock, DE} \ 
\textit{\IEEEauthorrefmark{4}Radboud University, Nijmegen, NL} \\ 
\{weixing,danstru\}@chalmers.se, jan-philipp.steghoefer@xitaso.com, regina.hebig@uni-rostock.de 
}}


\newcommand{\weixing}[1]{\textcolor{red}{#1}}
\newcommand{\regina}[1]{\textcolor{blue}{#1}}

\maketitle

\begin{abstract}
Metamodel-based DSL development in language workbenches like Xtext allows language engineers to focus more on metamodels and domain concepts rather than grammar details. 
However, the grammar generated from metamodels often requires manual modification, which can be tedious and time-consuming. Especially when it comes to rapid prototyping and language evolution, the grammar will be generated repeatedly, this means that language engineers need to repeat such manual modification back and forth.
Previous work introduced GrammarOptimizer, which automatically improves the generated grammar using optimization rules. However, the optimization rules need to be configured manually, which lacks user-friendliness and convenience. 
In this paper, we present our vision for and current progress towards a language workbench that integrates GrammarOptimizer's grammar optimization rules to support rapid prototyping and evolution of metamodel-based languages. It provides a visual configuration of optimization rules and a real-time preview of the effects of grammar optimization to address the limitations of GrammarOptimizer. Furthermore, it supports the inference of a grammar based on examples from model instances and offers a selection of language styles. These features aim to enhance the automation level of metamodel-based DSL development with Xtext and assist language engineers in iterative development and rapid prototyping. Our paper discusses the potential and applications of this language workbench, as well as how it fills the gaps in existing language workbenches.
\end{abstract}


\section{Introduction}
A \textbf{language workbench} is a powerful tool that facilitates the development of domain-specific languages (DSLs). It provides an integrated environment where language engineers can define, design, and implement languages to address specific application domains. 
There is a plethora of available language workbenches that cover a wide range of features, as identified and discussed in two previous surveys \cite{erdweg2015evaluating,iung2020systematic} of 10 and 14 workbenches, respectively.

We focus on a specific approach to the development of textual DSLs, where language engineers first create a metamodel and then automatically generate the grammar from the metamodel, followed by generating the editor from the grammar. We refer to this DSL development approach as \textbf{metamodel-based development}.
To our knowledge, the only available workbench supporting this approach is Xtext \cite{bettini2016implementing}.


Metamodel-based development entails several benefits for language engineers. 
First, language engineers focus on domain concepts and relationships at the beginning, rather than being burdened by syntax details from the start.
Second, at the beginning of the language creation process, it does not need to be clear whether the new language is text-based, graphical, or both~\cite{kleppe2007towards}.
In fact, following the philosophy of \textit{blended modeling}~\cite{ciccozzi2019blended}, there are good reasons to support multiple syntaxes, as different developers are likely to benefit from different syntax paradigms.
Thus, it is a sensible choice to start with the metamodel. Automatically generating a grammar from a metamodel helps the language engineer to save effort. 

However, there are limitations to the metamodel-based DSL development approach. The automatically generated grammar follows a fixed format that includes various grammar elements. Some of these elements are necessary, but the grammars typically becomes lengthy and relies on deep nesting, making it cumbersome to write code in the resulting DSL. As a result, the generated grammar often requires manual adjustments to be effectively used---we call this adjustment \textbf{grammar optimization}. 
Most importantly, the development of DSLs, just as the development of any software, requires iterations. To rapidly prototype a language, the metamodel, i.e. the conceptual side of the language, will be extended step by step, and so will the grammar. However, to keep the metamodel and grammar consistent, the grammar needs to be re-generated when the metamodel changes. This implies that manual grammar optimization needs to be re-applied, which is a cumbersome process. \emph{Therefore, we argue that there is today no language workbench that allows for metamodel-based DSL development, while also supporting rapid prototyping.}

In this paper, we present our vision for and progress toward developing such a language workbench.
Our envisioned language workbench will be based on the Xtext technology and will support the automatic optimization of grammars generated from metamodels, along with supporting rapid prototyping.

In previous work, we studied how to optimize generated grammars, leading to the development of a tool called GrammarOptimizer~\cite{zhang2023} that encompasses 54 optimization rules. Currently, it exists as a standalone grammar optimization tool, but we plan to integrate it into the envisioned language workbench, which will address several challenges associated with performing optimizations with GrammarOptimizer:
Before performing optimization operations with GrammarOptimizer, manual configuration of optimization rules is required, and the effects of the configuration are not visible in real-time during the configuration process. Moreover, the parametric approach to configuration requires new users to spend time learning how to use it. Although in specific scenarios, the ConfigGenerator we developed in~\cite{zhang2023sle} can automatically extract grammar optimization rule configurations, users still need to manually configure parameters in other scenarios.
To enhance usability and convenience, our envisioned language workbench will include a visual configuration of the grammar optimization rules and a real-time preview of the configuration effects. This preview will include the optimized grammar and example model instances that conform to the grammar. Additionally, we aim to provide optional language styles and the ability to infer grammar based on given model instances in the envisioned language workbench. Our goal is for the envisioned language workbench to facilitate the rapid prototyping of languages in iterative cycles of engineering, reducing manual work, shortening development time, and improving efficiency.

In the following, we focus on outlining the potential of the language workbench. We will provide insights into the specifics of our goals, discuss the architecture, features, and potential application scenarios of our language workbench, as well as address the limitations of existing language workbenches in these scenarios. 

\section{Motivating Example}
\label{sec:motivating_example}
We developed EATXT~\cite{EATXT}, a textual syntax for the architecture modeling language EAST-ADL with the help of our existing tool GrammarOptimizer.
We generated an Xtext grammar from the EAST-ADL metamodel, but the resulting grammar was not user-friendly and easy to use. In collaboration with our industrial partners, we aimed to make it more concise and user-friendly. We set specific modification goals to optimize the grammar, such as removing the keyword \texttt{`shortName'} from the attribute \texttt{shortName} in all the grammar rules and moving it from inside the outermost curly braces to before the opening one. Using GrammarOptimizer, we optimized a total of 2,046 lines of text and 233 grammar rules in the grammar. Despite the substantial number of optimization operations performed, we only needed to configure 31 lines of optimization rules. When developing the language, we needed to present the optimized effects to our industrial partners to ensure that the resulting grammar style met both parties' expectations. GrammarOptimizer saved us a significant amount of effort by supporting rapid prototyping. The generated grammar, the configuration of optimization rules, and the optimized grammar can be found in the supplementary material~\cite{zhang2023}.

However, we realized that using GrammarOptimizer leads to a substantial manual overhead. Configuring the rules requires invoking and parameterizing associated methods, which leads to complex code-based configurations. For instance, removing the keyword \texttt{`shortName'} requires the following configuration: ``\texttt{go.removeKeyword("EAPackage", "shortName", "shortName", null);}". The 1st two parameters specify the grammar rule and attribute where the keyword is to be removed, and the 3rd one is the keyword to be removed. The 4th parameter, here set to \textit{null}, can restrict the scope of the rule application, by specifying a context attribute determining the line on which changes will be performed. Additionally, during the configuration process, users lack visibility into the outcome of the optimization. The code-based configuration approach increases the risk of configuration errors, as language engineers may input incorrect parameter values, thus impacting the correctness and effectiveness of grammar optimization. These challenges have motivated us to develop a new language workbench to address these issues.

\section{Background and Related Work}
\label{sec:background}
\subsection{Existing Language Workbenches and Features}

The term "language workbench" was officially coined and popularized by Fowler in 2005~\cite{languageworkbench2005}. However, prior to that, tools and methods for language development had been evolving. The first version of MetaEdit was released in 1993~\cite{smolander1993metaedit+}, which served as the precursor to MetaEdit+. Today, MetaEdit+ has evolved into one of the most comprehensive language workbenches. In~\cite{erdweg2015evaluating}, Erdweg listed a total of 34 features covering various aspects, including notation, semantics, validation, testing, composability, editing mode, syntactic services, and semantic services. From their report, it can be observed that MetaEdit+ now covers the majority of these features, even supporting symbolic notation, which is not typically supported by many other language workbenches. Around 2003, JetBrains initiated an experimental project to validate the concept of language-oriented programming~\cite{pech2013jetbrains}, which eventually developed into the powerful language workbench JetBrains MPS. Today, JetBrains MPS encompasses almost all of the 34 features mentioned in~\cite{erdweg2015evaluating}, with the exceptions of graphical notation, free form, and live translation. In 2010, Spoofax was introduced as a language workbench specifically designed for creating textual languages~\cite{kats2010spoofax}. While it supports most of the aforementioned 34 features, it does not support DSL debugging, projectional editing, or quick fixes.

As mentioned before, to the best of our knowledge, these two papers cover the most comprehensive language workbenches currently available. 
Apart from Xtext, there is currently no other language workbench that supports metamodel-based DSL development. 
Although tools like GEMOC Studio enable language engineers to create metamodels, they do not support the direct generation of grammar from metamodels~\cite{combemale2017language}. Similarly, there are also metamodels in JetBrains MPS, but its underlying principles prevent it from supporting grammar generation from metamodels. The Grasland toolkit developed by Kleppe supports grammar generation from metamodels~\cite{kleppe2007towards}, but it is no longer publicly available today. In addition, David et al.~\cite{david2023blended} investigated existing modeling tools that support blended modeling, and the results showed that none of the tools supported the method proposed in this paper.

\subsection{Workflow of metamodel-based DSL development with Xtext}
\label{sec:traditional_workflow}
Eclipse Xtext is a framework for developing software languages, including modeling languages, and provides a complete infrastructure that includes tools, such as the parser, for the developed language~\cite{XtextHomepage}. Xtext offers two approaches for designing textual DSLs. One approach involves creating the grammar first and deriving the metamodel from it, while the other approach is the metamodel-based method mentioned in the previous section. Paige discusses and compares these two methods in~\cite{Paige. 2014}. This paper focuses on the latter method.

\begin{figure*}[tb]
  \centering
  \includegraphics[width=\linewidth]{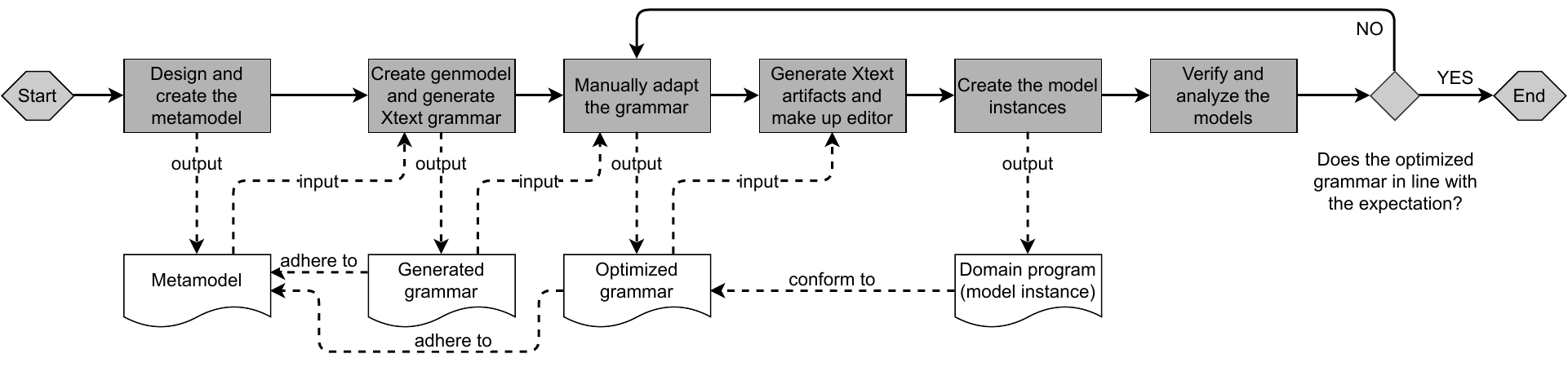}
  \caption{Workflow for developing a DSL in a metamodel-based approach using the Xtext framework.}
  \label{fig:traditional_workflow} 
\end{figure*}


Figure~\ref{fig:traditional_workflow} depicts the workflow of developing a DSL using the metamodel-based approach in Xtext. Initially, the language engineer creates a metamodel to represent domain concepts and relationships. Then, a generator model file, known as genmodel, is created based on the metamodel. Subsequently, the Xtext grammar is automatically generated from the genmodel. The generated grammar is manually modified, and the Xtext artifacts are generated to compose the DSL editor. Finally, the language engineer creates and verifies model instances within the DSL editor (i.e., the model instances are parsed correctly). If the language engineer finds that the grammar does not reach the desired state while verifying models in the DSL editor, she will iterate back to the manual modification of the grammar. This iterative process continues until the language engineer is satisfied with the grammar.



\subsection{GrammarOptimizer: Metamodel-Grammar Co-Evolution}
The mentioned grammar optimization refers to directly modifying the elements, e.g., the keywords, of a generated Xtext grammar. Before the development of GrammarOptimizer, no automated approaches existed for such modifications. In~\cite{zhang2023}, we analyzed the differences between the generated grammar and the desired grammar for seven DSLs and extracted 54 optimization rules based on the analysis. We developed the tool GrammarOptimizer to implement these optimization rules. GrammarOptimizer can run as a standalone Eclipse plugin. By configuring the optimization rules, it optimizes the generated grammar toward the desired grammar. The configuration for each optimization rule involves invoking its public method. When calling these methods, parameters are configured to specify the scope, target, content, and constraints of the optimization operation. For example, ``\texttt{go.addKeywordToAttr("Header", "appName", "pages", false);}" means adding the keyword \texttt{pages} after the attribute \texttt{appName} in the grammar rule \texttt{Header} (where \texttt{false} indicates that the keyword is added after the specified attribute).

The efforts invested in configuring GrammarOptimizer will pay off quickly when the language undergoes changes, such as during rapid prototyping iterations or continuous language evolution~\cite{zhang2023}. As an example, consider the evaluation section in~\cite{zhang2023} where we selected the EAST-ADL language. As part of the BUMBLE project~\cite{BUMBLE}, we developed a textual language, EATXT~\cite{EATXT}, for EAST-ADL. The development was divided into two stages~\cite{zhang2023}. In the first stage, we generated the grammar based on a simplified version of the EAST-ADL metamodel and configured 22 optimization rule settings to optimize the generated grammar towards the target appearance. The execution of these 22 configurations allowed us to avoid manual modifications in approximately 2,000 lines of text. In the second stage, we regenerated the grammar based on the full EAST-ADL metamodel, reusing the optimization rule settings from the simplified version, and optimized the generated grammar towards the target with only 10 modifications in the configurations. During the evolution of a metamodel-based language, GrammarOptimizer can save language engineers' efforts in adapting the grammar by reusing configurations with minimal modifications at each evolution step.

\section{Approach and Uniqueness}

\subsection{Visual Configuration of Optimization rules}



\begin{figure*}[h]
  \centering
  \includegraphics[width=\linewidth]{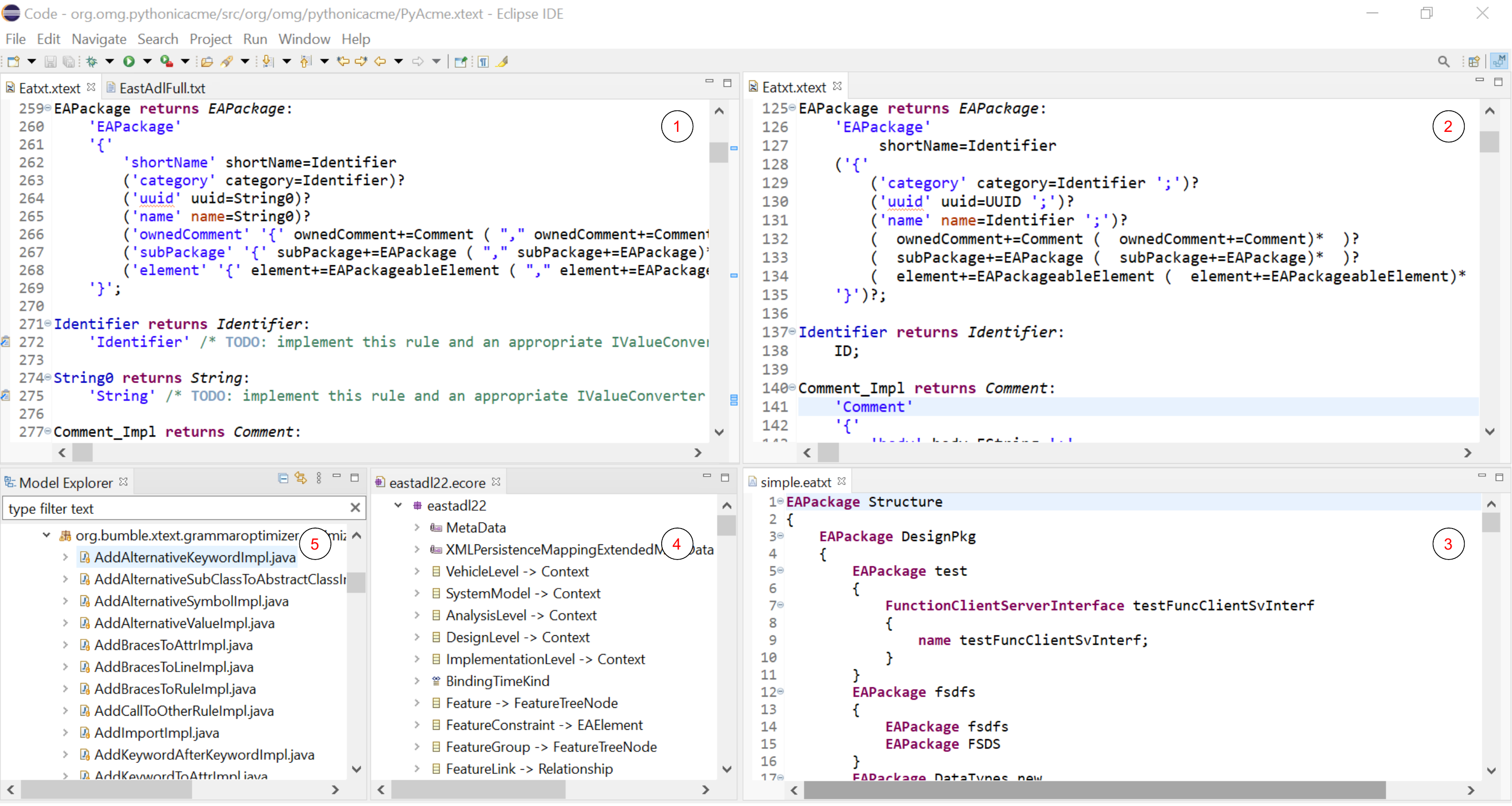}
  \caption{Visual work interface of the new language workbench.}
  \label{fig:visual_interface} 
\end{figure*}

Based on the challenges regarding the use of GrammarOptimizer for grammar optimization explained in Section~\ref{sec:motivating_example}, our envisioned language workbench will provide support for visual configuration. This means that language engineers will be able to configure optimization rules through a graphical user interface, integrated into the development environment they are familiar with. Our workbench will have a user interface with multiple windows, as depicted in Figure~\ref{fig:visual_interface}. Window \textcircled{1} will display the generated grammar and allow language engineers to select the grammar rules or elements they want to modify. When one or more grammar rules or elements are selected, the language engineer can configure the optimization rules in the standard user interface in Eclipse (\textit{properties} view). Configuring the optimization rules primarily involves selecting the optimization rules and parameterizing them, which will be done in that user interface. Parameterizing a selected optimization rule includes specifying the working scope of the optimization rule, the specific modifications to be performed, and any constraints associated with it. The working scope of an optimization rule refers to the places in the grammar that will be affected by the rule application, e.g., all lines of the specified grammar rule.

\subsection{Real-time Grammar Preview and Domain Programs}

In Section~\ref{sec:traditional_workflow}, we introduced the traditional workflow of metamodel-based DSL development. The problem is that each time we verify the modified grammar with model instances, we need to manually regenerate the editor and run it. Additionally, we optimize the grammar by configuring and performing the optimization rules, then the new issue is that the optimized grammar is not real-time shown when we configure the grammar.
The new language workbench addresses these challenges by providing real-time grammar previews and offering domain program examples. 
At the start, domain programs need to be imported or created by the language engineer. Later on, the language workbench will modify the domain programs according to the adjusting grammar.

As mentioned above, our workbench will consist of multiple windows. Window \textcircled{1} will display the generated grammar. Window \textcircled{2} will offer a preview of the optimized grammar.
When the language engineer configures the optimization rules, the grammar modified by these optimization rules can be seen in this window.
Window \textcircled{3} will show example model instances that conform to the optimized grammar. Through this window, language engineers can understand the impact of their optimization rule configurations on both grammar and model instances. 
Without such a capability, optimizing a grammar often involves iterative trial and error, where each iteration requires modifications, execution, and validation, sometimes even involving temporary model instances for experimentation. Language engineers spend considerable time and effort in this back-and-forth engineering process. However, this feature will reduce the time and effort of language engineers by supporting rapid prototype printing of languages, increasing their efficiency and productivity.

\begin{figure*}[tb]
  \centering
  \includegraphics[width=\linewidth]{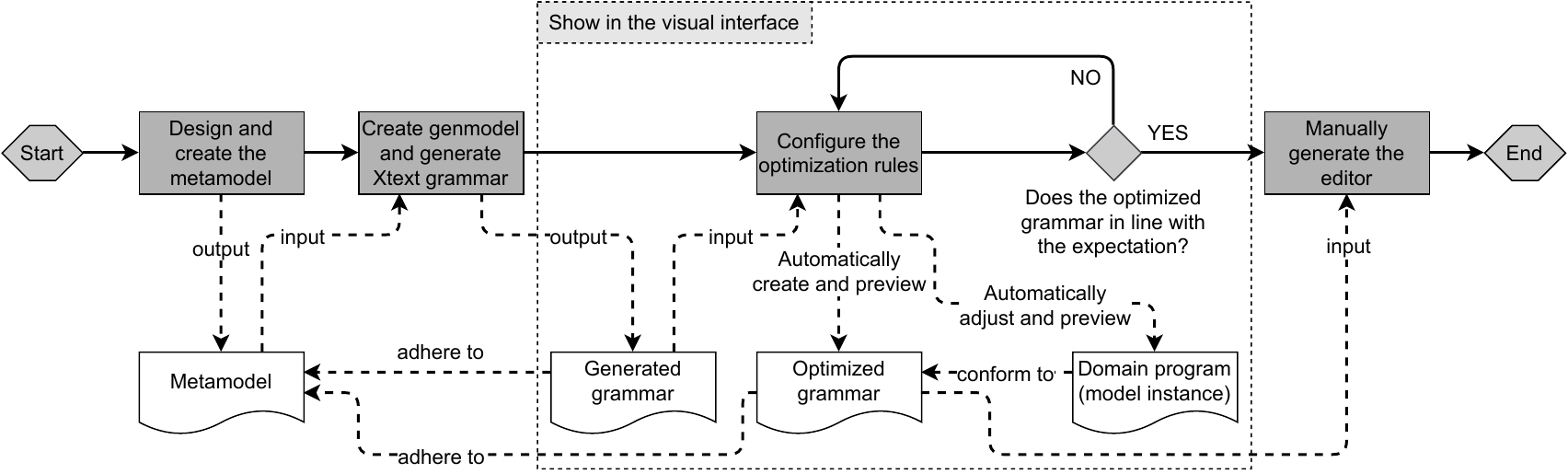}
  \caption{Workflow for developing a DSL with the envisioned language workbench}
  \label{fig:new_workflow} 
\end{figure*}

\subsection{Optional Language Styles}

We envision a scenario in which a language engineer is designing a new DSL where the appropriate style or target style is yet to be determined. Our envisioned language workbench will provide multiple optional language styles, such as Python style, Java style, C style, etc. A concrete example of this is a semi-automatic approach that we introduced in~\cite{pythonlikedsl2023} that can turn a DSL into a Python-like language.
Once the language workbench can optimize grammar, it will enable the swift transformation of a DSL into a language in a desired style. This transformation is achieved by executing corresponding optimization rule configurations.
Each language style option corresponds to a library that is composed of a set of pre-configured optimization rules. Users simply need to select the desired language style without any additional operations required. At the same time, we provide the ability for users to develop and contribute their style libraries. This means that users can write their style libraries with a set of configured rules and make them available for download and use by other users.

\subsection{Infer Grammar with Given Example Model Instance}

We envision the following scenario. To design a new language, a language engineer first tries to build the desired model instance to find out how the language should be designed. When she builds a trial model instance, she needs to design a DSL that can parse such a model instance.
Our envisioned language workbench will enable the derivation of grammar from a given model instance, upon which the language infrastructure is generated.

After importing the model instance file (i.e., the domain program), the language engineer selects the specific string in the given model instance on the interface and sets its type. For example, the language engineer selects a string whose content is \texttt{shortName} and sets its type to ``keyword''.
Based on this input, the workbench constructs corresponding grammar rules. Once all necessary grammar rules are constructed, the workbench generates Xtext artifacts from the grammar, forming the complete set of infrastructure required for the editor. Ultimately, the imported model instance program will be parsed and displayed by the editor.

This approach allows language engineers to rapidly create a DSL tailored to specific model instances without investing significant development time in a full-fledged DSL, making it a suitable solution for one-time parsing needs.


\section{Workflow of Language Engineers}
\label{sec:new_workflow}
The envisioned language workbench aims to save the effort of language engineers and improve the efficiency and productivity of designing languages by allowing rapid prototyping. In contrast to the workflow depicted in Figure~\ref{fig:traditional_workflow}, where language engineers manually modify the generated grammar, the envisioned language workbench allows engineers to configure optimization rules instead. While configuring these rules, engineers can preview the optimized grammar and model instances that conform to the grammar in the visual interface. By observing the real-time preview, language engineers can assess whether the optimization rule configurations can change the generated grammar to the desired one. This new workflow is shown in Figure~\ref{fig:new_workflow}.

In a language evolution or rapid prototyping scenario, the metamodel undergoes changes resulting in a new version. For such scenarios, the workflow in our language workbench would involve re-generating the grammar and then reusing the configuration from the first version. This initial configuration, which optimized the grammar of the first version, serves as a starting point for optimizing the grammar of the second version. This way previously made decisions about the syntax can be re-applied. Language engineers would then adjust the configuration to account for the differences in the concepts of the metamodel introduced by the evolution process. By iteratively adapting and refining the configuration, the optimized grammar for the new language version can be achieved.

\section{Conclusion and Future Work}

In this paper, we envision a new language workbench designed to support rapid prototyping and evolution of languages to improve the workflow of language engineers leveraging Xtext for metamodel-based DSL development.
This language workbench will integrate the grammar optimization capabilities of the GrammarOptimizer, i.e., the optimization rules, and incorporate the features for visually configuring the optimization rules and real-time preview of the optimization effects. There will also be optional library styles in the language workbench, such as the Python style, where each style consists of a set of optimization rule configurations. These library styles can be customized and extended by users. Additionally, we envision implementing the capability of example-based grammar inference within this new language workbench. By integrating these capabilities and features, the automation level and user-friendliness of the language workbench will be enhanced.

We foresee several future directions for our work. Firstly, we plan to develop the feature of visual optimization rule configurations for the language workbench, which primarily involves interface development. Secondly, we aim to implement a real-time preview of grammar optimization, which involves addressing two technical challenges, i.e., 1) parsing multiple grammar files simultaneously and 2) parsing examples from model instances. Thirdly, we will introduce various library styles to offer a broader range of choices, which will involve experimenting with multiple language styles. Finally, we will explore how to infer matching grammars based on given model instance programs.

\bibliographystyle{IEEEtran}
\bibliography{mle}

\begin{thebibliography}{10}
\providecommand{\url}[1]{#1}
\csname url@samestyle\endcsname
\providecommand{\newblock}{\relax}
\providecommand{\bibinfo}[2]{#2}
\providecommand{\BIBentrySTDinterwordspacing}{\spaceskip=0pt\relax}
\providecommand{\BIBentryALTinterwordstretchfactor}{4}
\providecommand{\BIBentryALTinterwordspacing}{\spaceskip=\fontdimen2\font plus
\BIBentryALTinterwordstretchfactor\fontdimen3\font minus
  \fontdimen4\font\relax}
\providecommand{\BIBforeignlanguage}[2]{{%
\expandafter\ifx\csname l@#1\endcsname\relax
\typeout{** WARNING: IEEEtran.bst: No hyphenation pattern has been}%
\typeout{** loaded for the language `#1'. Using the pattern for}%
\typeout{** the default language instead.}%
\else
\language=\csname l@#1\endcsname
\fi
#2}}
\providecommand{\BIBdecl}{\relax}
\BIBdecl

\bibitem{erdweg2015evaluating}
S.~Erdweg, T.~Van Der~Storm, M.~V{\"o}lter, L.~Tratt, R.~Bosman, W.~R. Cook,
  A.~Gerritsen, A.~Hulshout, S.~Kelly, A.~Loh \emph{et~al.}, ``Evaluating and
  comparing language workbenches: Existing results and benchmarks for the
  future,'' \emph{Computer Languages, Systems \& Structures}, vol.~44, pp.
  24--47, 2015.

\bibitem{iung2020systematic}
A.~Iung, J.~Carbonell, L.~Marchezan, E.~Rodrigues, M.~Bernardino, F.~P. Basso,
  and B.~Medeiros, ``Systematic mapping study on domain-specific language
  development tools,'' \emph{Empirical Software Engineering}, vol.~25, pp.
  4205--4249, 2020.

\bibitem{bettini2016implementing}
L.~Bettini, \emph{Implementing domain-specific languages with Xtext and
  Xtend}.\hskip 1em plus 0.5em minus 0.4em\relax Packt Publishing Ltd, 2016.

\bibitem{kleppe2007towards}
A.~Kleppe, ``Towards the generation of a text-based ide from a language
  metamodel,'' in \emph{Model Driven Architecture-Foundations and Applications:
  Third European Conference, ECMDA-FA 2007, Haifa, Israel, June 11-15, 2007,
  Proccedings 3}.\hskip 1em plus 0.5em minus 0.4em\relax Springer, 2007, pp.
  114--129.

\bibitem{ciccozzi2019blended}
F.~Ciccozzi, M.~Tichy, H.~Vangheluwe, and D.~Weyns, ``Blended modelling-what,
  why and how,'' in \emph{2019 ACM/IEEE 22nd International Conference on Model
  Driven Engineering Languages and Systems Companion (MODELS-C)}.\hskip 1em
  plus 0.5em minus 0.4em\relax IEEE, 2019, pp. 425--430.

\bibitem{zhang2023}
W.~Zhang, J.~Holtmann, R.~Hebig, and J.-P. Steghöfer, ``Meta-model-based
  language evolution and rapid prototyping with automate grammar
  optimization,'' Preprint available at SSRN:
  \url{http://dx.doi.org/10.2139/ssrn.4379232}, 2023.

\bibitem{zhang2023sle}
W.~Zhang, R.~Hebig, D.~Strüber, and J.-P. Steghöfer, ``Automated extraction
  of grammar optimization rule configurations for metamodel-grammar
  co-evolution,'' in \emph{16th ACM SIGPLAN International Conference on
  Software Language Engineering (SLE'23)}, 2023, (in press).

\bibitem{EATXT}
J.~Holtmann, J.-P. Steghöfer, and W.~Zhang, ``Exploiting meta-model structures
  in the generation of {Xtext} editors,'' in \emph{11th Intl. Conf. on
  Model-Based Software and Systems Engineering (MODELSWARD)}, 2023, pp.
  218--225, (in press).

\bibitem{languageworkbench2005}
{Martin Fowler}, ``Language workbenches: The killer-app for domain specific
  languages?'' 2005,
  \url{https://martinfowler.com/articles/languageWorkbench.html}. Last accessed
  Jul 2023.

\bibitem{smolander1993metaedit+}
K.~Smolander, ``Metaedit+ protocols and standard operations for processing
  goprr information structures: the application programmer's interface,''
  \emph{Internal Technical Document, MetaPHOR project, Univ. of
  Jyv{\"a}skyl{\"a}, Jyv{\"a}skyl{\"a}, Finland}, 1993.

\bibitem{pech2013jetbrains}
V.~Pech, A.~Shatalin, and M.~Voelter, ``Jetbrains mps as a tool for extending
  java,'' in \emph{Proceedings of the 2013 International Conference on
  Principles and Practices of Programming on the Java Platform: Virtual
  Machines, Languages, and Tools}, 2013, pp. 165--168.

\bibitem{kats2010spoofax}
L.~C. Kats and E.~Visser, ``The spoofax language workbench: rules for
  declarative specification of languages and ides,'' in \emph{Proceedings of
  the ACM international conference on Object oriented programming systems
  languages and applications}, 2010, pp. 444--463.

\bibitem{combemale2017language}
B.~Combemale, O.~Barais, and A.~Wortmann, ``Language engineering with the gemoc
  studio,'' in \emph{2017 IEEE International Conference on Software
  Architecture Workshops (ICSAW)}.\hskip 1em plus 0.5em minus 0.4em\relax IEEE,
  2017, pp. 189--191.

\bibitem{david2023blended}
I.~David, M.~Latifaj, J.~Pietron, W.~Zhang, F.~Ciccozzi, I.~Malavolta,
  A.~Raschke, J.-P. Stegh{\"o}fer, and R.~Hebig, ``Blended modeling in
  commercial and open-source model-driven software engineering tools: A
  systematic study,'' \emph{Software and Systems Modeling}, vol.~22, no.~1, pp.
  415--447, 2023.

\bibitem{XtextHomepage}
{Eclipse Foundation}, ``Xtext homepage,'' 2022,
  \url{https://www.eclipse.org/Xtext/}. Last accessed Nov 2022.

\bibitem{BUMBLE}
{ITEA}, ``Bumble project,'' 2023, \url{https://itea4.org/project/bumble.html}.
  Last accessed Jul 2023.

\bibitem{pythonlikedsl2023}
W.~Zhang, R.~Hebig, J.-P. Steghöfer, and J.~Holtmann, ``Creating python-style
  domain specific languages: A semi-automated approach and intermediate
  results,'' in \emph{11th Intl. Conf. on Model-Based Software and Systems
  Engineering (MODELSWARD)}, 2023, pp. 210--217, (in press).

\end{thebibliography}

\vspace{12pt}
\color{red}

\end{document}